\documentclass[12pt]{article}
\usepackage{epsfig}

\usepackage{amssymb}
\usepackage{amsmath}
\usepackage{amsfonts}

\def\be{\begin{equation}}
\def\ee{\end{equation}}
\def\ba{\begin{array}{c}}
\def\ea{\end{array}}

\def\ben{$$}
\def\een{$$}
\begin{document}

\titlepage

 \begin{center}{\Large \bf

Rectifiable ${\cal PT}-$symmetric Quantum Toboggans with Two Branch
Points

 }\end{center}

\vspace{5mm}

 \begin{center}
Miloslav Znojil

\vspace{3mm}

Nuclear Physics Institute ASCR\footnote{e-mail: znojil@ujf.cas.cz },
250 68 \v{R}e\v{z}, Czech Republic

\end{center}

\vspace{5mm}

\section*{Abstract}

Certain complex-contour (a.k.a. quantum-toboggan) generalizations of
Schr\"{o}dinger's bound-state problem are reviewed and studied in
detail. Our key message is that the practical numerical solution of
these atypical eigenvalue problems may perceivably be facilitated
via an appropriate complex change of variables which maps their
multi-sheeted complex domain of definition to a suitable
single-sheeted complex plane.

 \vspace{9mm}

\noindent
 PACS 03.65.Ge


 \begin{center}

\end{center}

\newpage

\section{Introduction \label{s1} }
%

One-dimensional Schr\"{o}dinger equation for bound states
 \be
 -\frac{\hbar^2}{2m}\,\frac{d^2}{dx^2} \,\psi_n (x) + V(x)
  \,\psi_n (x)= E_n \,\psi_n (x)\,, \ \ \ \
  \ \ \ \
 \psi_n (x)\in \mathbb{L}^2(\mathbb{R})
   \label{SE}
 \ee
belongs among the most friendly phenomenological models in quantum
mechanics \cite{Fluegge}. For virtually all of the reasonable
phenomenological confining potentials $V(x)$ the numerical treatment
of this eigenvalue problem remains entirely routine.

During certain recent numerical experiments \cite{experiments} it
became clear that many standard (e.g., Runge-Kutta \cite{RK})
computational methods may still encounter new challenges when one
follows the advice by Bender and Turbiner \cite{BT}, by Buslaev and
Grecchi \cite{BG}, by Bender et al \cite{BBjmp} or by Znojil
\cite{tobog} and when one replaces the most common real line of
coordinates $x \in \mathbb{R}$ in ordinary differential
Eq.~(\ref{SE}) by some less trivial complex contour of $x \in {\cal
C}(s)$ which may be conveniently parametrized, whenever necessary,
by a suitable real pseudocoordinate $s \in \mathbb{R}$,
 \be
 -\frac{\hbar^2}{2m}\,\frac{d^2}{dx^2} \,\psi_n (x) + V(x)
  \,\psi_n (x)= E_n \,\psi_n (x)\,, \ \ \ \
  \ \ \ \
 \psi_n (x)\in \mathbb{L}^2({\cal C})
   \label{SEPTIK}\,.
 \ee
Temporarily, the scepticism has been suppressed by Weideman
\cite{Weideman} who showed that many standard numerical algorithms
may be reconfirmed to lead to reliable results even for many
specific analytic samples of {\em complex} interactions $V(x)$
giving real spectra via Eq.~(\ref{SEPTIK}).

Unfortunately, the scepticism reemerged when we proposed, in
Ref.~\cite{tobog}, to study the so called quantum toboggans
characterized by the relaxation of the most common tacit assumption
that the above-mentioned integration contours ${\cal C}(s)$ must
always lie just inside a {\em single} complex plane ${\cal R}_0$
equipped by suitable cuts. Subsequently, the reemergence of certain
numerical difficulties accompanying the evaluation of the spectra of
quantum toboggans has been reported by B\'{\i}la \cite{Bila} and by
Wessels \cite{Wessels}. Their empirical detection of the presence of
instabilities in their numerical results may be recognized as one of
the key motivations of our present considerations.

\section{Illustrative tobogganic Schr\"{o}dinger equations}


\begin{figure}[t]                     
\begin{center}                         
\epsfig{file=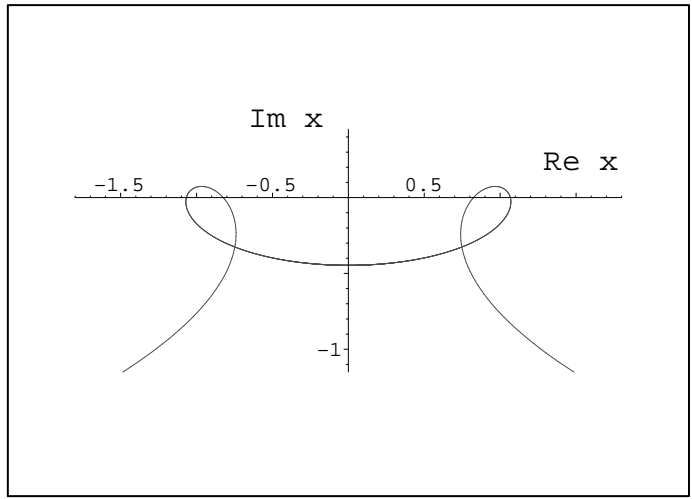,angle=270,width=0.6\textwidth}
\end{center}                         
\vspace{-2mm} \caption{The central segment of the typical ${\cal
PT}-$symmetric double-circle tobogganic curve of $x \in {\cal
C}^{(LR)}(s)$ with winding parameter $\kappa=3$ in
Eq.~(\ref{change}). This curve is obtained as the image of the
straight line of $z \in {\cal C}^{(0)}(s)$ at $\varepsilon=0.250$.
 \label{obr1}}
\end{figure}

\subsection{Assumptions}


\begin{figure}[t]                     
\begin{center}                         
\epsfig{file=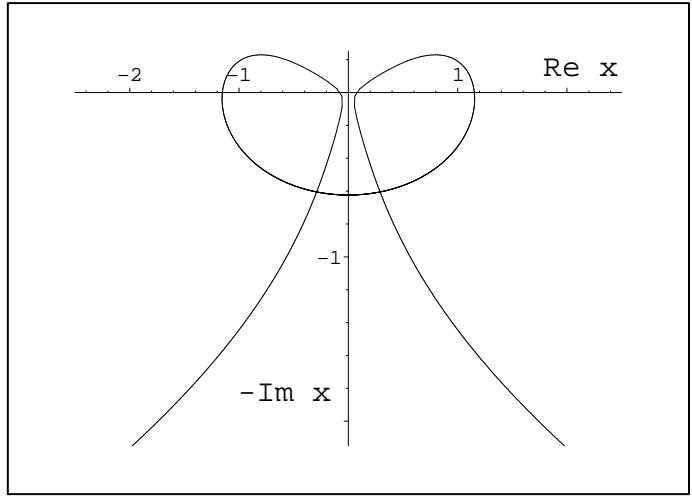,angle=270,width=0.6\textwidth}
\end{center}                         
\vspace{-2mm} \caption{An alternative version of the double-circle
curve of Figure \ref{obr1} obtained  at the ``almost maximal"
$\varepsilon=\varepsilon^{(critical)}-0.0005$ (note that
$\varepsilon^{(critical)} \sim 0.34062502$).
 \label{obr2}}
\end{figure}

\begin{figure}[t]                     
\begin{center}                         
\epsfig{file=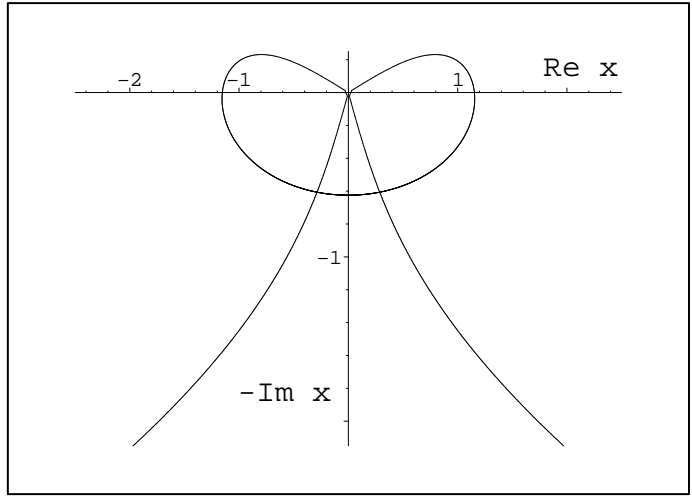,angle=270,width=0.6\textwidth}
\end{center}                         
\vspace{-2mm} \caption{The extreme version of the double-circle
curve ${\cal C}^{(LR)}(s)$ at $\varepsilon  \lessapprox
\varepsilon^{(critical)}$.
 \label{obr3}}
\end{figure}

\begin{figure}[t]                     
\begin{center}                         
\epsfig{file=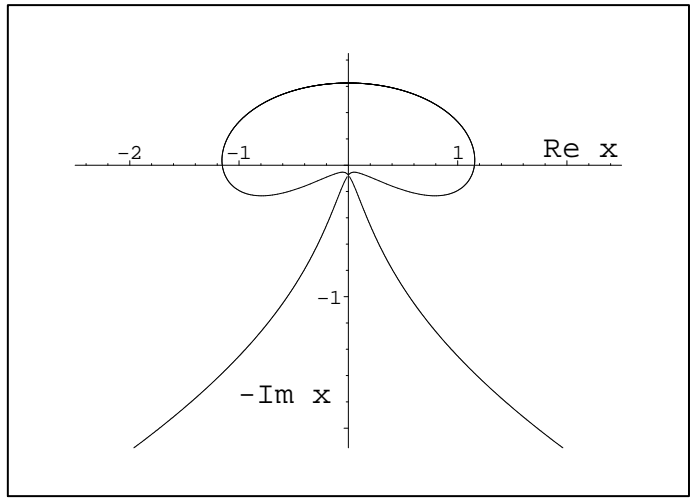,angle=270,width=0.6\textwidth}
\end{center}                         
\vspace{-2mm} \caption{The change of topology at
  $\varepsilon  \gtrapprox \varepsilon^{(critical)}$ when
  Eq.~(\ref{change}) starts giving
  the single-circle tobogganic curves ${\cal C}^{(RL)}(s)$ at
 $\kappa=3$.
 \label{obr4}}
\end{figure}

Whenever the complex integration contour ${\cal C}(s)$ used in
Eq.~(\ref{SEPTIK}) becomes topologically nontrivial (cf.
Figures~\ref{obr1} -- \ref{obr4} for illustration), it may be
interpreted as connecting {\em several sheets} of the Riemann
surface ${\cal R}^{(multisheeted)}$ supporting the general solution
$\psi^{(general)}(x)$ of the underlying complex ordinary
differential equation. It is well known that these solutions
$\psi^{(general)}(x)$ are non-unique (i.e., two-parametric - cf.
\cite{Bila}). From the point of view of physics this means that they
may be restricted by some suitable (i.e., typically, asymptotic
\cite{BT,BG}) boundary conditions (cf. also Ref.~\cite{tobog}). In
what follows we shall assume that

\begin{itemize}

 \item [(A1)]
these general solutions $\psi^{(general)}(x)$ live on unbounded
contours called ``tobogganic", with the name coined and with the
details explained in Ref.~\cite{tobog};

 \item [(A2)]
our particular choice of the tobogganic contours
 \ben
 {\cal C}(s)={\cal C}^{(tobogganic)}(s)\in {\cal
R}^{(multisheeted)}
 \een
will be specified by certain multiindex $\varrho$ so that  ${\cal
C}^{(tobogganic)}(s)\equiv {\cal C}^{(\varrho)}(s)$;

 \item [(A3)]
for the sake of brevity our attention may be restricted to the
tobogganic models where the multiindices $\varrho$ are nontrivial
but still not too complicated. For this reason we shall  study just
the subclass of the tobogganic models
 \be
 -\frac{\hbar^2}{2m}\,\frac{d^2}{dx^2} \,\psi_n (x) + V^{(2)}_{(j)}(x)
  \,\psi_n (x)= E_n \,\psi_n (x)\,, \ \ \ \
  \ \ \ \
 \psi_n (x)\in \mathbb{L}^2({\cal C}^{(\varrho)})
   \label{SEPTIKtob}\,
 \ee
containing, typically, potentials
 \be
 V^{(2)}_{(1)}(x) = V_{(HO)}(x)= x^2+
 \left [
 \frac{F}{(x-1)^2}+
 \frac{F}{(x+1)^2}
 \right ]\,,\ \ \ \ F \gg 1
 \label{HO}
 \ee
or
 \be
  V^{(2)}_{(2)}(x)=V_{(ICO)}(x)= {\rm i}x^3+
 \left [
 \frac{G}{(x-1)^2}+
 \frac{G}{(x+1)^2}
 \right ]\,,\ \ \ \ G \gg 1\,
 \label{HOb}
 \ee
with two strong singularities inducing branch points in the wave
functions.

\end{itemize}

 \noindent
In this manner  we shall have to deal with the two branch points
$x^{(BP)}_{(\pm)}=\pm 1$ in $\psi^{(general)}(x)$. In the language
of mathematics the obvious topological structure of the
corresponding multi-sheeted Riemann surface ${\cal
R}^{(multisheeted)}$ will be ``punctured" at $x^{(BP)}_{(\pm)}=\pm
1$.  In the vicinity of these two ``spikes" we shall assume the
generic, ``logarithmic" \cite{web} structure of ${\cal
R}^{(multisheeted)}$.


\subsection{Winding descriptors $\varrho$}

The multiindex $\varrho$ will be called ``winding descriptor" in
what follows. It will be used here in the form introduced in
Ref.~\cite{bitobog} where each curve ${\cal C}^{(\varrho)}(s)$ has
been assumed moving from its ``left asymptotics" (where $s \ll -1$)
to a point which lies below one of the branch points
$x^{(BP)}_{(\pm)}=\pm 1$. During the further increase of $s$ one
simply selects one of the following four alternative possibilities:

\begin{itemize}

 \item
one moves counterclockwise around the left branch point
$x^{(BP)}_{(-)}$ (this move is represented by the first letter $L$
in the ``word" $\varrho$),

 \item
one moves counterclockwise around the right branch point
$x^{(BP)}_{(+)}$ (this move is represented by letter $R$),

 \item
one moves clockwise around the left branch point $x^{(BP)}_{(-)}$
(this move is represented by letter $Q$ or symbol $L^{-1}\equiv Q$),

 \item
one moves clockwise around the right branch point $x^{(BP)}_{(+)}$
(this move is represented by letter $P$ or symbol $R^{-1}\equiv P$).

\end{itemize}

 \noindent
In this manner we may compose the moves and characterize each
contour by a word $\varrho$ composed of the sequence of letters
selected from the four-letter alphabet $R,L,Q$ and $P$. Once we add
the requirement of ${\cal PT}-$symmetry (i.e., of a left-right
symmetry of contours) we arrive at the sequence of eligible words
$\varrho$ of even length $2N$.

At $N=0$ we may assign the empty symbol $\varrho=\emptyset$ or
$\varrho=0$ to the one-parametric family of the straight lines of
Ref.~\cite{BG},
 \be
 {\cal C}^{(0)}(s)\ \equiv\  s - {\rm i}\,\varepsilon\,,
 \ \ \ \ \varepsilon > 0.
 \label{nontobo}
 \ee
Thus, one encounters precisely four possible arrangements of the
descriptor, viz.,
 \be
 \varrho \in \left \{LR\,, L^{-1}R^{-1}\,, RL\,,R^{-1}L^{-1}
 \right \}\,,\ \ \ \ \ N=1\,
 \label{ill1}
 \ee
in the first nontrivial case. In the more complicated cases where $N
> 1$ it makes sense to re-express the requirement of  ${\cal PT}-$symmetry
in the form of the string-decomposition $\varrho= \Omega \bigcup
\Omega^T$ where the superscript $^T$ marks an {\em ad hoc}
transposition, i.e., the reverse reading accompanied by the $L
\leftrightarrow R$ interchange of symbols. Thus, besides the
illustrative Eq.~(\ref{ill1}) we may immediately complement the
first nontrivial list
 \ben
 \Omega \in \left \{L\,, L^{-1}\,, R\,,R^{-1}
 \right \}\,,\ \ \ \ \ N=1\,,
 \een
by its $N=2$ descendant
 \be
 \left \{
 LL,
 LR, RL, RR, L^{-1}R, R^{-1}L,
  LR^{-1}\,, RL^{-1}\,,
  L^{-1}L^{-1}, L^{-1}R^{-1}, R^{-1}L^{-1},R^{-1}R^{-1}
 \right \}
 \label{ill2}
 \ee
etc. The four ``missing" words $ L L^{-1}\,, L^{-1}L\,,R R^{-1}$ and
$R^{-1}R$ had to be omitted as trivial here because they cancel each
other when interpreted as windings \cite{bitobog}.

\section{Rectifications}


\begin{figure}[t]                     
\begin{center}                         
\epsfig{file=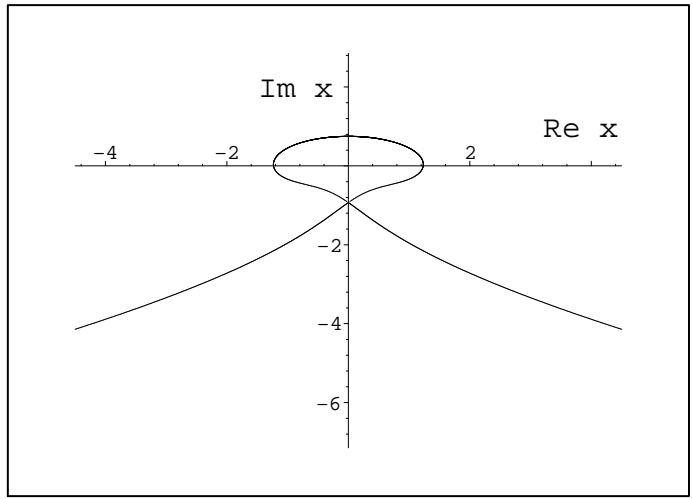,angle=270,width=0.6\textwidth}
\end{center}                         
\vspace{-2mm} \caption{The  fully developed version of the
 single-circle tobogganic curve ${\cal C}^{(RL)}(s)$ obtained at
 $\kappa=3$ and
  $\varepsilon =0.400 $.
 \label{obr5}}
\end{figure}

\subsection{Formula}

The core of our present message lies in the idea that the
non-tobogganic straight lines (\ref{nontobo}) may be mapped on their
specific (called ``rectifiable") tobogganic descendants. For this
purpose one may use the following closed-form recipe of
Ref.~\cite{bitobog},
 \be
 {\cal M}:
 \left (z\ \in \ {\cal C}^{(0)}(s)
 \right ) \ \to \
 \left (x\ \in \ {\cal C}^{(\varrho)}(s)
 \right )
 \ee
where one defines
 \be
 x = -{\rm i}\,\sqrt{(1-z^2)^\kappa - 1}\,.
 \label{change}
 \ee
This formula guarantees the ${\cal PT}$ symmetry of the resulting
contour as well as the stability of the position of our pair of the
branch points. Another consequence of this choice is that the
negative imaginary axis of $z=-{\rm i}|z|$ is mapped upon
itself.

Some purely numerical features of the mapping (\ref{change}) may be
also checked via the freely available software of
Ref.~\cite{Novotny}. On this empirical basis we shall demand that
the exponent $\kappa$ will be chosen here as an odd positive
integer, $\kappa=2M+1$, $M=1,2,\ldots$. In this case the asymptotics
of the resulting nontrivial tobogganic contours (with $M \neq 0$)
will still parallel the $\kappa=1$ real line ${\cal C}^{(0)}(s)$ in
the leading-order approximation.

\subsection{The sequences of critical points \label{cripo}}

The inspection of Figures \ref{obr2} and \ref{obr3} and their
comparison with Figures \ref{obr4} and \ref{obr5} reveals that one
should expect the emergence of sudden changes of the winding
descriptors $\varrho$ during a smooth variation of the shift
$\varepsilon>0$ of the initial straight line of $z$ introduced via
Eq.~(\ref{nontobo}). Formally we may set $\varrho=
\varrho(\varepsilon)$ and mark the set of the corresponding points
of changes of $\varrho(\varepsilon)$ by the sub- and superscript in
$\varepsilon^{(critical)}_{j}$.

The quantitative analysis of these critical points is not difficult
since it gets perceivably simplified by the graphical insight gained
via Figures \ref{obr2} -- \ref{obr4} and via their appropriately
selected more complicated descendants. Trial and error constructions
enable us to formulate (and, subsequently, to prove) the very useful
hypothesis that the transition between different descriptors
$\varrho(\varepsilon)$ always proceeds via the same mechanism. Its
essence is characterized by the confluence and ``flip" of the curve
at any $j=1,2,\ldots, M$ in $\varepsilon=
\varepsilon^{(critical)}_{j}$. At this point certain two branches of
the curve ${\cal C}^{(\varrho)}(s)$  touch and reconnect in the
manner sampled by the transition from Figure \ref{obr2} to Figure
\ref{obr4}.

The key characteristics of this flip is that it takes place in the
origin so that we can determine the point $x^{(critical)}_{j}=0$
which carries the obvious geometric meaning mediated by the complex
mapping (\ref{change}). Thus, the vanishing $x^{(critical)}_{j}=0$
is to be perceived as an image of some doublet of
$z=z^{(critical)}_{j}$ or, due to the left-right symmetry of the
picture, as an image of a symmetric pair of the pseudocoordinates
$s^{(critical)}_{j} =\pm \left | s^{(critical)}_{j}\right |$.

At any $\kappa =2M+1$ the latter observations reduce
Eq.~(\ref{nontobo}) to elementary relation
 \be
 1 = \left \{
 1+[{\rm i}(s - {\rm i}\,\varepsilon)]^2\right \}^\kappa\,
 \label{tarov}
 \ee
which may be analyzed in the equivalent form of the following $2M+1$
independent relations
 \be
 e^{2\pi{\rm i}\, m /(2M+1)} =
 1+({\rm i}s +\varepsilon)^2=1+\varepsilon^2-s^2
 +2\,{\rm i}s\,\varepsilon
 \,.
 \label{tarovbe}
 \ee
These relations numbered by $ m = 0, \pm 1,\ldots, M$ may further be
simplified via the two known elementary trigonometric real and
non-negative constants $A$ and $B$ such that
 \ben
 \left [1-e^{2\pi{\rm i}\, m /(2M+1)} \right ] = A \pm {\rm i}\,B\,.
 \een
In terms of these constants we separate Eq.~(\ref{tarovbe}) into it
real and imaginary parts yielding the pair of relations
 \be
 s^2-\varepsilon^2-A=0\,,\ \ \ \ \
 2\,s\,\varepsilon = B\,.
 \ee
As long as $\varepsilon>0$ we may restrict our attention to the
non-negative $s$ and eliminate $s=B/(2\,\varepsilon)$. The remaining
quadratic equation
 \ben
 B^2/(2\,\varepsilon)^2-\varepsilon^2-A=0\,
 \een
finally leads to the following unique solution of the problem,
 \be
 \varepsilon=\frac{1}{\sqrt{2}}\,\sqrt{-A+\sqrt{A^2+B^2}}\,.
 \ee
This formula perfectly confirms the validity and precision of our
illustrative graphical constructions.

%
%
%
%
%

%
%
%
%
%
%
%
%
%
%
%
%
%
%
%
%
%
%
%
%
%
%
%
%
%
%
%
%
%

\section{Samples of countours of complex coordinates}

For the most elementary toboggans characterized by the single
branching point the winding descriptor $\varrho$ becomes trivial
because it is being formed by the words in a one-letter alphabet.
This means that all the information about windings degenerates just
to the length of the word $\varrho$ represented by an (arbitrary)
integer \cite{ostobo}. Obviously, these models would be too trivial
from our present point of view.

In an opposite direction one could also contemplate tobogganic
models where a larger number of branch points would have to be taken
into account. An interesting series of exactly solvable models of
this form may be found, e.g., in Ref. \cite{Anjana}. Naturally, the
study of all of these far reaching generalizations would still
proceed along the lines which are tested here on the first
nontrivial family characterized by the presence of the mere two
branch points in $\psi(x)$.

From the pedagogical point of view the merits of the
two-branch-point scenario involve not only the simplicity of the
formulae (cf., e.g., Eq.~(\ref{change}) in preceding section) but
also the feasibility and transparency of the graphical presentation
of the integration contours ${\cal C}^{(\varrho)}$ of the tobogganic
Schr\"{o}dinger equations. This assertion may easily be supported by
a few explicit illustrative pictures.


\begin{figure}[t]                     
\begin{center}                         
\epsfig{file=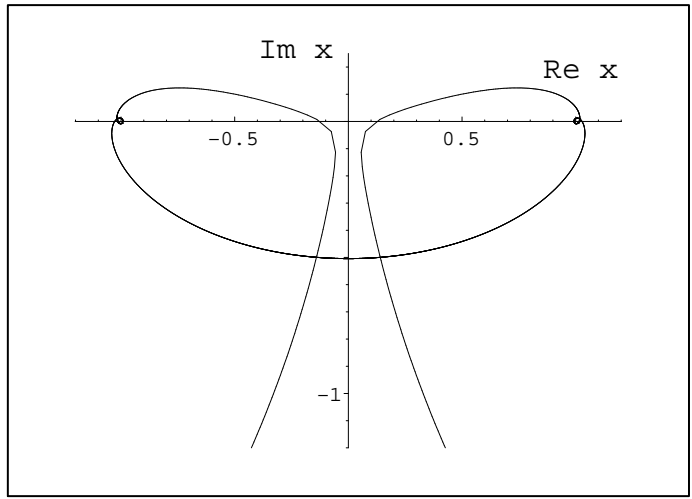,angle=270,width=0.6\textwidth}
\end{center}                         
\vspace{-2mm} \caption{The quadruple-circle tobogganic curve of $x
\in {\cal C}^{(LLRR)}(s)$. With winding parameter $\kappa=5$ in
Eq.~(\ref{change}) this sample is obtained at
$\varepsilon=\varepsilon^{(critical)}_{1}-0.0005$, i.e., just
slightly below the first critical value of
$\varepsilon^{(critical)}_{1} \sim 0.21574990$.
 \label{obr6}}
\end{figure}

\subsection{Rectifiable tobogganic contours with
$\kappa = 3$}

The change of variables (\ref{change}) generating the rectifiable
tobogganic Schr\"{o}dinger equations must be implemented with due
care because the knot-shaped curves ${\cal C}^{(\varrho)}(s)$ may
happen to run quite close to the points of singularities at certain
values of $s$. This is well illustrated by Figure \ref{obr1} or,
even better, by Figure \ref{obr6}. At the same time all our Figures
clearly show that one can control the proximity to the singularities
by means of the choice of the shift $\varepsilon$ of the
(conventionally chosen) straight line of the auxiliary variable $z
\in {\cal C}^{(0)}$ given by Eq.~(\ref{nontobo}).

Once we fix the distance $\varepsilon$ of the complex line $ {\cal
C}^{(0)}$ from the real line $\mathbb{R}$ we may still vary the odd
integers $\kappa$. {\it Vice versa}, even at the smallest $\kappa=3$
the recipe enables us to generate certain mutually non-equivalent
tobogganic contours ${\cal C}^{(\varrho)}(s)$ in the
$\varepsilon-$dependent manner. This confirms the existence of
discontinuities. Their emergence and form are best illustrated by
the pair of Figures \ref{obr3} and \ref{obr4}.


\begin{figure}[thb]                     
\begin{center}                         
\epsfig{file=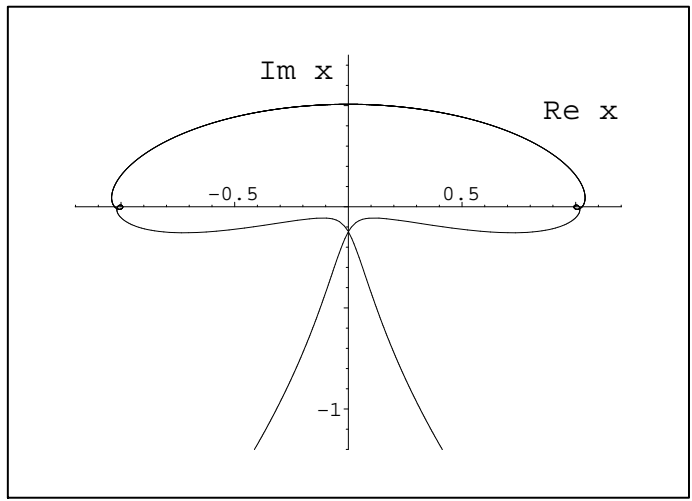,angle=270,width=0.6\textwidth}
\end{center}                         
\vspace{-2mm} \caption{The topologically different, triple-circle
curve ${\cal C}^{(RRLL)}(s)$ obtained at $\kappa=5$ and
$\varepsilon=\varepsilon^{(critical)}_{1}+0.0005$.
 \label{obr7}}
\end{figure}

We may conclude that in general one has to deal here with the very
high sensitivity of the results to the precision of the numerical
input or to the precision of computer arithmetics. This confirms the
expectations expressed in our older paper \cite{bitobog} where we
emphasized that the descriptor $\varrho$ is not necessarily easilly
inferred from a nontrivial, detailed analysis of the mapping ${\cal
M}$.

\subsection{Rectifiable tobogganic contours with
$\kappa \geq 5$}

Once we select the next odd integer $\kappa=5$ in Eq.~(\ref{change})
the study of the knot-shaped structure of the resulting integration
contours ${\cal C}^{(\varrho)}(s)$ becomes even more involved
because in the generic case sampled by Figure \ref{obr6} the size of
the internal loops proves unexpectedly small in comparison. As a
consequence, their very existence may in principle escape our
attention. Thus, one might even mistakenly perceive the curve of
Figure \ref{obr6}  as an inessential deformation  of the curves in
Figures \ref{obr1} or \ref{obr2}.


\begin{figure}[t]                     
\begin{center}                         
\epsfig{file=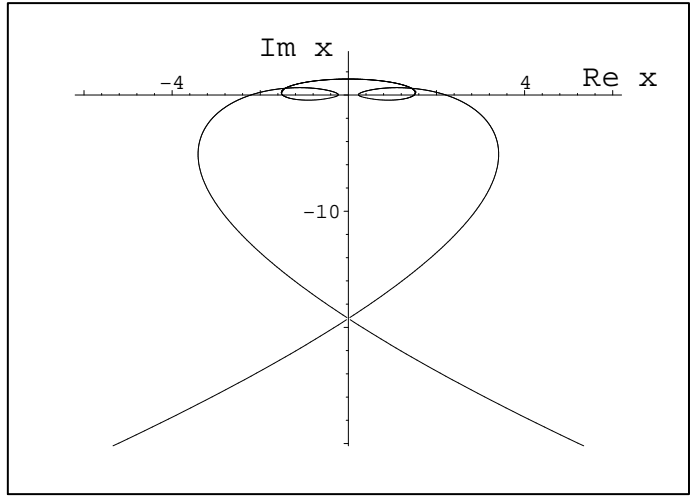,angle=270,width=0.6\textwidth}
\end{center}                         
\vspace{-2mm} \caption{The other extreme triple-circled $\kappa=5$
curve ${\cal C}^{(RRLL)}(s)$ as emerging at
$\varepsilon=\varepsilon^{(critical)}_{2}-0.005$, i.e., close to the
second boundary $\varepsilon^{(critical)}_{2} \sim 0.49223343$.
 \label{obr8}}
\end{figure}
%


\begin{figure}[t]                     
\begin{center}                         
\epsfig{file=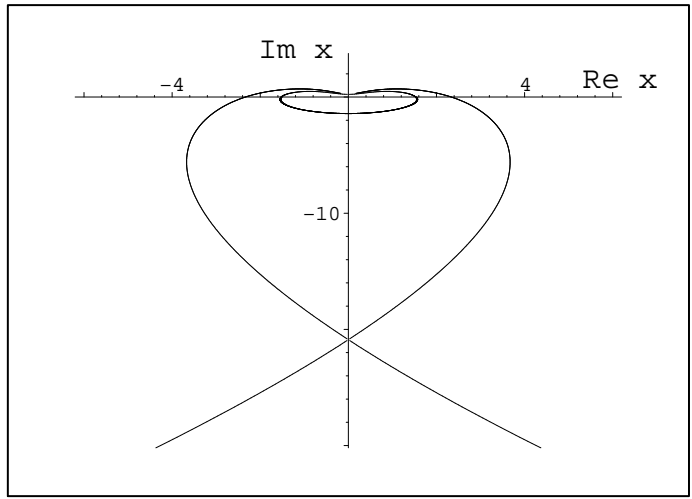,angle=270,width=0.6\textwidth}
\end{center}                         
\vspace{-2mm} \caption{The twice-circling tobogganic $\kappa=5$
curve ${\cal C}^{(RLRL)}(s)$ as emerging slightly above the second
critical shift-parameter, viz., at
$\varepsilon=\varepsilon^{(critical)}_{2}+0.005$.
 \label{obr9}}
\end{figure}

Naturally, not all of the features of our toboganic integration
contours will change during transition from $\kappa=3$ to
$\kappa=5$. In particular, the partial parallelism between Figures
\ref{obr2} and \ref{obr6} survives as the similar global-shape
partial parallelism between Figures \ref{obr4} (where $\kappa = 3$)
and \ref{obr7} (where $\kappa = 5$). Moreover, a certain local-shape
partial parallelism may be also found between Figure \ref{obr2}
(where the two upwards-oriented loops almost touch at $\kappa = 3$)
and Figure \ref{obr8} (where the two downwards-oriented ``inner"
loops almost touch at $\kappa = 5$). The latter parallels seem to
sample certain more general mechanism since Figure \ref{obr4} also
finds its replica inside the upper part of Figure \ref{obr9}, etc.
Obviously, the next-step transition from $\kappa=5$ to $\kappa=7$
(etc.) may be also expected to proceed along similar lines.

For the computer-assisted drawing of the graphical representation of
the curves ${\cal C}^{(\varrho)}$ the formulae of paragraph
\ref{cripo} should be recalled as the source of the most useful
information about the critical parameters. The extended-precision
values of the underlying coordinates  of the points of instability
are needed in such an application. Their $M \leq 6$ sample is listed
here in Table~1.


\begin{table}[h]
\caption{Transition parameters for $\kappa=2M+1$ with $M = 1, 2,
\ldots,6$} \label{pexp4}
\begin{center}
\begin{tabular}{||c|c|c||c||c|c||}
\hline \hline
 M&&
 {\rm } & $\varepsilon^{(critical)}_{(m)}$
 &{\rm pseudocoordinate} &{\rm angle}
 \\
 &
 m&B&$\left [ \right.${\rm critical shift in} $\left.{\cal C}^{(0)}(s) \right ]$&
 $\left |s^{(critical)}_{(m)}\right |$&$\varphi^{(critical)}_{(m)}$\\
 \hline
  \hline
        1
                                {} & 1
                                                   {} & 0.8660
                     {} & 0.34062501931660664017
                     {} & 1.2712
                    {} & 0.2618
                               \\ 2
                                {} & 1

                     {} & 0.9510

                     {} & 0.49223342986833679823

                     {} & 0.96606

                    {} & 0.4712

                               \\

                                {} & 2

                     {} & 0.5878

                     {} & 0.21574989943840034163

                     {} & 1.3622

                    {} & 0.1571

                               \\ 3

                                {} & 1

                     {} & 0.7818

                     {} & 0.49560936234793313854

                     {} & 0.78876

                    {} & 0.5610

                               \\

                                {} & 2

                     {} & 0.9749

                     {} & 0.41300244005317039597

                     {} & 1.1803

                    {} & 0.3366

                               \\

                                {} & 3

                     {} & 0.4339

                     {} & 0.15634410200136762402

                     {} & 1.3876

                    {} & 0.1122


                               \\ 4

                                {} & 1

                     {} & 0.6428

                     {} & 0.47438630343334929661

                     {} & 0.67749

                    {} & 0.6109


                               \\

                                {} & 2

                     {} & 0.9848

                     {} & 0.47917814904271720218

                     {} & 1.0276

                    {} & 0.4363

                               \\

                                {} & 3

                     {} & 0.8660

                     {} & 0.34062501931660664017

                     {} & 1.2712

                    {} & 0.2618

                               \\

                                {} & 4

                     {} & 0.3420

                     {} & 0.12231697600600608108

                     {} & 1.3981

                    {} & 0.08727

                               \\ 5

                                {} & 1

                     {} & 0.5406

                     {} & 0.44984366535166445772

                     {} & 0.60092

                    {} & 0.6426

                               \\

                                {} & 2

                     {} & 0.9096

                     {} & 0.49834558687374848153

                     {} & 0.91265

                    {} & 0.4998


                               \\

                                {} & 3

                     {} & 0.9898

                     {} & 0.42964189183273983152

                     {} & 1.1519

                    {} & 0.3570

                               \\

                                {} & 4

                     {} & 0.7557

                     {} & 0.28670826353957054964

                     {} & 1.3180

                    {} & 0.2142

                               \\

                                {} & 5

                     {} & 0.2817

                     {} & 0.10037407570525388131

                     {} & 1.4034

                    {} & 0.071400


                               \\ 6

                                {} & 1

                     {} & 0.4647

                     {} & 0.42666576745054519911

                     {} & 0.54460

                    {} & 0.6646

                               \\

                                {} & 2

                     {} & 0.8230

                     {} & 0.49875399287559237235

                     {} & 0.82504

                    {} & 0.5437

                               \\

                                {} & 3

                     {} & 0.9927

                     {} & 0.47264256935707423545

                     {} & 1.0502

                    {} & 0.4229

                               \\

                                {} & 4

                     {} & 0.9350

                     {} & 0.38168235795277279438

                     {} & 1.2249

                    {} & 0.3021

                               \\

                                {} & 5

                     {} & 0.6631

                     {} & 0.24649719795540125795

                     {} & 1.3451

                    {} & 0.1812

                               \\

                                {} & 6

                     {} & 0.2393

                    {} & 0.085076232785825555735

                     {} & 1.4065

                    {} & 0.06042

 \\
  \hline
  \hline
\end{tabular}
 \label{tabone}
\end{center}
\end{table}

On this basis we may summarize that at a generic $\kappa$ the
variation (i.e., in all of our examples, the growth) of the shift
$\varepsilon$ makes certain subspirals of contours ${\cal
C}^{(\varrho})$ larger and moving closer and closer to each other.
In this context our Table~1 could, in principle, serve as a certain
systematic guide towards a less intuitive classification of our
present graphical pictures characterizing transitions between
different winding descriptors ${\varrho}$ and, hence, between the
topologically non-equivalent rectifiable tobogganic contours ${\cal
C}^{(\varrho)}$. During such phase-transition-like processes
\cite{BT} the value of $\varepsilon$ crosses a critical point beyond
which the asymptotics of the contours are changing. As a
consequence, also the spectra of the underlying tobogganic quantum
bound-state Hamiltonians will get, in general, changed \cite{topol}.

\section{Conclusions}

We confirmed the viability of an innovated, ``tobogganic" version of
${\cal PT}-$symmetric Quantum Mechanics of bound states in models
where the general solutions of the underlying ordinary differential
Schr\"{o}dinger equation exhibit two branch-point singularities
located, conveniently, at $x^{(BP)}=\pm 1$.

In particular we clarified that many topologically complicated
complex integrations contours which spiral around the branch points
$x^{(BP)}$ in various ways may be rectified. This means that one can
apply an elementary change of variables $z(s) \to x(s)$ and replace
the complicated original tobogganic quantum bound-state problem by
an equivalent simplified differential equation defined along the
straight line of complex pseudocoordinates $z=s-{\rm i}\varepsilon$.

In detail a few illustrative rectifications have been  described
where we succeeded in an assignment of the {\em different} winding
descriptors $\varrho$ to the tobogganic contours controlled solely
by the variation of the ``initial" complex shift $\varepsilon$. An
interesting supplementary result of our present considerations may
be also seen in the constructive demonstration of feasibility of an
explicit description of these transitions between topologically
non-equivalent quantum toboggans characterized by non-equivalent
winding descriptors $\varrho$. Still, the full understanding of
these structures remains to be an open problem recommended to a
deeper analysis in the nearest future.

In summary we have to emphasize that our present
rectification-mediated reconstruction of the
ordinary-differential-equation representation of quantum toboggans
could be perceived as an important step towards their rigorous
mathematical analysis and, in particular, towards the extension of
the existing rigorous proofs of the reality/observability of the
energy spectra to these promising innovative phenomenological
models.

\subsection*{Acknowledgements}

The support by the Institutional Research Plan AV0Z10480505 and by
the M\v{S}SMT ``Doppler Institute" project LC06002 is acknowledged.


\newpage

\begin{thebibliography}{00}

\bibitem{Fluegge}
%
Fl\"{u}gge, S.: Practical Quantum Mechanics I, II.  Berlin,
Springer, 1971.


\bibitem{experiments}
%
Znojil, M.: Experiments in PT-symmetric quantum mechanics. Czech. J.
Phys. Vol. 54 (2004), p. 151 -- 156 (quant-ph/0309100v2).

\bibitem{RK}
%
Znojil, M.: One-dimensional Schr¨odinger equation and its ``exact"
representation on a discrete lattice. Phys. Lett. Vol. A 223 (1996).
p. 411-416.
%
%
%
%

\bibitem{BT}
%
Bender, C. M., Turbiner, A. V.: Analytic Continuation of Eigenvalue
Problems. Phys. Lett. Vol. A 173 (1993), p. 442-445.

\bibitem{BG}
%
Buslaev, V.,  Grechi, V.:  Equivalence of unstable anharmonic
oiscllators and double wells. J. Phys. A: Math. Gen. Vol. 26 (1993),
p. 5541 - 5549.


\bibitem{BBjmp}
%
Bender, C. M., Boettcher, S.:  Real spectra in non-Hermitian
Hamiltonians having PT symmetry.
 Phys. Rev. Lett. Vol. { 80} (1998), p. 5243 - 5246;

Bender, C. M., Boettcher, S., Meisinger, P. N.: PT-symmetric quantum
mechanics.
  J. Math. Phys.
Vol. 40 (1999), p. 2201.

\bibitem{tobog}
%
Znojil, M.: PT-symmetric quantum toboggans. Phys. Lett. Vol. A 342
(2005), p. 36-47.

\bibitem{Weideman}
%
Weideman, J. A. C., Spectral differentiation matrices for the
numerical solutions of Schr\"{o}dinger equation. J. Phys. A: Math.
Gen. Vol. 39 (2006), p. 10229 - 10238.

\bibitem{Bila}
%
B\'{\i}la, H.: Non-Hermitian Operators in Quantum Physics (PhD
thesis supervised by M. Znojil). Prague, Charles University, 2008;

B\'{\i}la, H.: Pramana - J. Phys. Vol. 73 (2009), p. 307 -- 314.


\bibitem{Wessels}
%
Wessels, G. J. C.: A numerical and analytical investigation into
non-Hermitian Hamiltonians (Master-degree thesis supervised by H. B.
Geyer  and J. A. C. Weideman). Stellenbosch, University of
Stellenbosch, 2008.

\bibitem{web}
%
see, e.g., "branch point" in http://eom.springer.de.

\bibitem{bitobog}
%
Znojil, M.: Quantum toboggans with two branch points. Phys. Lett.
Vol. A 372 (2008) p. 584 - 590 (arXiv: 0708.0087).

\bibitem{Novotny}
%
Novotn\'{y}, J.:
http://demonstrations.wolfram.com/TheQuantumTobogganicPaths.


\bibitem{ostobo}
%
Znojil, M.: Spiked potentials and quantum toboggans. J. Phys. A:
Math. Gen. Vol. 39 (2006), p. 13325-13336 (quant-ph/0606166v2);

Znojil, M.: Identification of observables in quantum toboggans. J.
Phys. A: Math. Theor. Vol. 41 (2008), p. 215304 (arXiv:0803.0403);

Znojil, M.: Quantum toboggans: models exhibiting a multisheeted PT
symmetry. J. Phys.: Conference Series, Vol. 128 (2008), p. 012046;

Znojil, M., Geyer, H. B.: Sturm-Schroedinger equations: formula for
metric. Pramana  J. Phys. Vol. 73 (2009), p. 299 - 306
(arXiv:0904.2293).

\bibitem{Anjana}
%
Sinha, A., Roy, P.: Generation of exactly solvable non-Hermitian
potentials with real
  energies.
Czech. J. Phys. Vol. {54} (2004), p. 129--138.

\bibitem{topol}
Znojil, M.: Topology-controlled spectra of imaginary cubic
oscillators in the large-L approach. Phys. Lett. Vol. A 374 (2010),
p. 807–812 
(arXiv:0912.1176v1).


%
%
%
%
%
%
%
%
%
%
%
%




%
%
%
%
%
%
%
%
%
%
%
%
%
%
%
%
%
%
%

%
%
%
%
%
%
%
%
%
%
%
%




\end{thebibliography}
\end{document}